%% file: 4dqg.tex
\documentstyle[12pt]{article}
\newif\ifFigureInclude \FigureIncludefalse
\FigureIncludetrue
\ifFigureInclude\input{epsf}
\else{\typeout{(Figure won't be included)}}\fi

\newcommand{\fig}[1]{Fig.~\ref{fig:#1}}

\catcode`@=11 
\def\lsim{\mathrel{\mathpalette\@versim<}}
\def\gsim{\mathrel{\mathpalette\@versim>}}
\def\@versim#1#2{\lower0.2ex\vbox{\baselineskip\z@skip
\lineskip\z@skip\lineskiplimit\z@\ialign{$\m@th#1
\hfil##\hfil$\crcr#2\crcr\sim\crcr}}}

\newcommand{\beq}{\begin{equation}}
\newcommand{\eeq}{\end{equation}}
\def\bsub{\begin{mathletters}}
\def\esub{\end{mathletters}}

\newcommand{\beqa}{\begin{eqnarray}}
\newcommand{\eeqa}{\end{eqnarray}}

\newcommand {\av}{\mbox{{\scriptsize av}}}
\newcommand {\tot}{\mbox{{\scriptsize tot}}}
\newcommand {\lat}{\mbox{{\scriptsize lat}}}

\newcommand {\dd}{\mbox{d}}

\def\lamA{5\cdot 10^{-3}}
\def\lamB{5\cdot 10^{-4}}
\def\lamC{5\cdot 10^{-6}}

\begin{document}
\setlength{\oddsidemargin}{0cm}
\setlength{\baselineskip}{7mm} 

\begin{titlepage}
 \renewcommand{\thefootnote}{\fnsymbol{footnote}}
    \begin{normalsize}
     \begin{flushright}
                 KEK-TH-439\\
                 May 1994\\
~~\\
~~\\
     \end{flushright}
    \end{normalsize}
    \begin{Large}
       \vspace{1cm}
       \begin{center}
         {\Large 
Singular Vertices in the Strong Coupling Phase 
of Four--Dimensional Simplicial Gravity} \\
       \end{center}
    \end{Large}

  \vspace{10mm}

\begin{center}
           Tomohiro H{\sc otta}$^{1)}$\footnote
           {
e-mail address : hotta@danjuro.phys.s.u-tokyo.ac.jp,
{}~JSPS research fellow.},
           Taku I{\sc zubuchi}$^{1)}$\footnote
           {
e-mail address : izubuchi@danjuro.phys.s.u-tokyo.ac.jp,
{}~JSPS research fellow.}{\sc and},
           Jun N{\sc ishimura}$^{2)}$\footnote
           {e-mail address : nisimura@theory.kek.jp,
{}~JSPS research fellow.}\\
      \vspace{1cm}
        $^{1)}$ {\it Department of Physics, 
University of Tokyo ,} \\
                 {\it Hongo, Bunkyo-ku, Tokyo 113, Japan}\\
        $^{2)}$ {\it National Laboratory for High Energy Physics (KEK),}\\
               {\it Tsukuba, Ibaraki 305, Japan}\\
\end{center}

\vspace{15mm}

\begin{abstract}
\noindent 
\setlength{\baselineskip}{5mm}
We study four--dimensional simplicial gravity through numerical simulation 
with special attention to 
the existence of singular vertices, in the strong coupling phase, 
that are shared by abnormally large 
numbers of four--simplices. 
We attempt to cure this disease by 
adding a term to the action which suppresses such singular vertices. 
For a sufficiently large coefficient of the additional term, however, 
the phase transition disappears and the system is observed to be 
always in the branched polymer phase for any gravitational constant. 
\end{abstract}

\end{titlepage}
\vfil\eject



\setcounter{footnote}{0}

Nowadays, due to precise measurements of the gauge coupling constants 
at LEP, it is usual to consider the three interactions which are
described quite well by the Standard Model to be unified into 
a Grand Unified Theory at about the energy scale of $10^{16}$ GeV. 
On the other hand, it is well known that the gravitational interaction 
between particles becomes nonnegligible at the Planck scale, which is 
around $10^{19}$ GeV. 
These two energy scales are remarkably close to each other
considering the ambiguity involved in the derivation of the above values. 
The most natural interpretation of this fact is 
that, around such a high energy scale, all the 
four interactions, including gravity, are unified. 
There are two possibilities for such unified theories at present.  
One is string theory and the other is a unification 
within ordinary field theory including gravity. 

The problem we encounter when we try to formulate quantum gravity within 
ordinary field theory in four dimensions is 
that we cannot renormalize it perturbatively. 
In the path integral formalism, quantization of the 
geometry is performed by integrating over the metric field. 
If we use lattice regularization, which enables a nonperturbative 
study, general coordinate invariance is not 
manifest and whether it is restored in the continuum limit is a crucial 
problem. 
There are two kinds of lattice regularization of quantum gravity; 
namely dynamical triangulation and Regge calculus, of which the former 
is believed to restore general coordinate 
invariance in the continuum limit. 
It has been solved exactly in two dimensions 
\cite{DT} and its 
continuum limit is shown to reproduce Liouville theory 
\cite{DDK}, in which general coordinate invariance has been 
treated carefully. 
There is also a handwaving argument for the restoration of general coordinate 
invariance in the continuum limit of dynamical triangulation \cite{handwaving}. 

Although four--dimensional dynamical triangulation 
seems to be difficult to solve analytically, there is no potential barrier 
in studying it through numerical simulations.
Employing the Einstein--Hilbert action as the lattice action and sweeping
the gravitational constant, 
it has been discovered that the system undergoes a second order 
phase transition \cite{4DQG}, which suggests the possibility 
of taking a continuum limit.
The geometrical features of each phase have the following 
characteristics. 
In the weak coupling phase, the configurations are made of relatively small 
baby universes connected through pinches. 
This phase is called the branched polymer phase. 
In the strong coupling phase, on the other hand, the configuration has a 
mother universe with very high connectivity, 
which seems to be due to the existence of vertices shared 
by very large numbers of four--simplices. 
This phase is called the crumpled phase.

In the continuum theory, it is well--known that Euclidean Gravity 
may have a problem related to the unboundedness of the Einstein term 
\cite{instability}. 
Decomposing the metric into the conformal mode and the transversal mode, 
one finds that the unboundedness comes from the fact that we can make 
the Einstein term arbitrarily negative
by choosing rapidly varying conformal mode.
This problem of the unboundedness of the Einstein term is, therefore, 
called `the conformal mode instability'. 
Since the coefficient of the Einstein term is proportional 
to the inverse of the gravitational constant, 
it is natural for us to relate this problem to 
the branched polymer structure observed in the weak coupling phase. 

In this respect, the strong coupling phase seems to be more promising 
than the weak coupling phase. 
At least, there is a mother universe, although it might
be plagued with singular vertices shared by abnormally large numbers of 
four--simplices. 
Let us call the number of four--simplices sharing a vertex the vertex order. 
In this paper, we measure the vertex order distribution with high statistics. 
We observe two singular vertices which show themselves as a sharp peak. 
Moreover, we find that the peak moves towards a larger value of vertex order linearly as we increase 
the system size. 
Considering this fact to be a potential obstacle 
to taking a continuum limit, 
we add a term in the action which suppresses such singular 
vertices. 
For a sufficiently large coefficient of the additional term, however, 
we find that the phase transition disappears and 
that the system is always in the branched polymer phase. 

~

Let us first give a general outlook on this kind of approach.
When we regularize four--dimensional quantum gravity 
with dynamical triangulation the integration over the metric 
is replaced with the random summation over 
all the four--dimensional simplicial manifolds. 
It is this point of the formalism that is essential in 
recovering general coordinate invariance in the continuum limit, 
as can be seen from the handwaving argument in 
Ref. \cite{handwaving}. 
Note that the lattice action can, therefore, 
be chosen arbitrarily as long as it is local 
without being restricted by general coordinate invariance.
We have to search for a sound second order phase transition, 
where we can take a sensible continuum limit, by changing the lattice action 
in various ways. 
Considering universality in quantum gravity 
as well as in ordinary field theories, 
we expect that the continuum theory, if it exists at all, 
does not depend on the way we construct it. 

It is natural for us to start with the Einstein--Hilbert action 
\begin{equation}
  S = \int \dd^4  x \sqrt{g} \left( \Lambda - \frac{1}{G} R \right)
\label{eq:EHaction}
\end{equation}
as a first trial, 
where $\Lambda$ is the cosmological constant and $G$ is the
gravitational constant.
Let us denote the number of $i$-simplices in a simplicial manifold by 
$N_i$. 
One can easily find that, for a simplicial manifold, 
\beqa
\int \dd^4 x \sqrt{g}  &=& c N_4 \\
\int \dd^4 x \sqrt{g} R  &=& 2 \pi N_2 - 10 \alpha N_4 ,
\eeqa
where $c$ is the volume of each four--simplex and 
$\alpha$ is the angle between two faces of a four--simplex, 
which is equal to $\arccos \left(\frac{1}{4}\right)$. 
Therefore the Einstein--Hilbert action (\ref{eq:EHaction}) can be 
expressed in terms of lattice variables as
\begin{equation}
  S_{\lat} = \kappa_4 N_4 - \kappa_2 N_2 ,
  \label{eq:lataction}
\end{equation}
where $\kappa_4$ and $\kappa_2$ are related to $\Lambda$ and $G$ through,
\begin{eqnarray}
  \kappa_4 & = & c \Lambda + \frac{10 \alpha}{G} \\
  \kappa_2 & = & \frac{2 \pi}{G}. 
\label{eq:kappa2G}
\end{eqnarray}
The $N_i$'s $(i = 0, 1, \cdots, 4)$ satisfy the following three relations. 
\beqa
N_0-N_1+N_2-N_3+N_4 &=&\chi  \\
2N_1-3N_2+4N_3-5N_4 &=&0  \\
5N_4 &=& 2N_3,
\eeqa
where $\chi$ is the Euler number, which is two for the spherical topology.
Therefore only two of the $N_i$'s are independent, which means 
that, actually, the lattice action 
(\ref{eq:lataction}) is the most general one
that can be written as a linear combination of $N_i$'s.

We consider an ensemble with a fixed $N_4$ and 
with spherical topology. 
There are well established methods for generating 
such an ensemble through numerical simulations, 
and the technical details of our simulation shall be given elsewhere \cite{HIN}
except for a few comments. 
The initial configuration is taken to be the surface of a 
five--simplex and 
we use the so--called ($p$,$q$)--moves \cite{Gross92}
to update a configuration. 
Since the topology is not changed by the moves, 
the spherical topology of the initial configuration is 
maintained throughout the run. 
Our code is written for arbitrary dimension following Ref. \cite{Catt94}.



Let us turn to the results of our simulation.
We first look at the second order phase transition, which can be seen 
through thermodynamic quantities such as the average curvature per 
unit volume : 
%
\beqa
R_{\av} &=& \frac{R_{\tot}}{N_4} ~~~~~~~~~~~~~~~~~~~~~~~~~~~~~~
(R_{\tot}= \int \dd^4  x \sqrt{g} R)  \\
       &=&  2 \pi \frac{N_2}{N_4} - 10 \alpha .
\eeqa
\label{eq:Rave}
\fig{fig1} shows our results for $\langle R_{\av} \rangle$
at various $\kappa_2$'s.
%
%
In contrast to the three--dimensional case \cite{3DQG}, 
no hysteresis has been observed. 
Also, one sees that the size dependence of the data changes abruptly 
at $\kappa_2 = 1.2 \sim 1.3$. On the right there is little size 
dependence, whereas on the left, the curve goes lower and lower as
we increase the system size. 
The derivative of the average curvature gives the susceptibility 
\begin{equation}
\chi_R = 2 \pi  \frac{\partial\langle R_{\av}\rangle}{\partial \kappa_2} 
     = \frac{\langle R_{\tot}^2 \rangle - 
       \langle R_{\tot}\rangle^2}{N_4} ,
\end{equation}
which represents the fluctuation of the total curvature. 
\fig{fig2} shows the result.
As is expected from \fig{fig1}, 
the susceptibility has a peak 
around $\kappa_2 = 1.2 \sim 1.3$, which grows higher as the
system size is increased. 
%
%
This implies that the correlation length of the local curvature diverges 
at the critical point \cite{twopoint}, where we may hope to take a 
continuum limit. 
These results are in reasonable agreement with the previous data 
in Refs. \cite{4DQG,Brug93a}.
Since $\kappa_2$ corresponds to the inverse of the gravitational constant, 
as is seen from (\ref{eq:kappa2G}), we call the large $\kappa_2$ phase as 
the weak coupling phase 
and the small $\kappa_2$ phase as the strong coupling phase.

However, as we mentioned earlier, 
the properties in both phases seem to be rather strange. 
In particular, considering the possibility of taking a continuum 
limit from the strong coupling phase, 
the singular vertices that are shared by abnormally large 
numbers of four--simplices 
might be an obstacle. 
We would like to study this phenomenon in detail.

Let us look at the vertex order distribution, which can be defined as 
\beq
\rho(n) = \frac{1}{N_0} \langle \sum _{v} \delta_{o(v),n} \rangle,
\eeq
where $o(v)$ is the order of the vertex $v$. 
This quantity is measured every 100 sweeps and averaged 
over 100 configurations. 
In order to reduce the fluctuations of the distribution, 
we smear the data over bins of size 10.

We first note that the average vertex order $\overline{o(v)}$ can be given as
\beq
\overline{o(v)} 
= \frac{1}{N_0} \sum_v o(v) = \frac{5 N_4}{N_0}.
\eeq
In the last equality, we used the relation
\begin{equation}
  \sum_v o(v) = 5 N_4,
\end{equation}
which comes from the fact that each four-simplex has five vertices. 
Thus one finds that the average curvature per unit volume 
$R_{\av}$ can be written in terms of $\overline{o(v)}$ as
\beqa
R_{\av} &=&  2 \pi \frac{N_2}{N_4} - 10 \alpha\\
       &=&  20 \pi \overline{o(v)}^{-1} - (10 \alpha -4 \pi)
             - \frac{8 \pi}{N_4} \\
       &\sim& 20\pi \overline{o(v)}^{-1} - 0.614~~~~~~~~~(N_4 \gg 1),
\eeqa
where in the second equality we used the relation 
\begin{equation}
  N_2 = 2 (N_0 + N_4 -2 ).
\label{eq:N2N0}
\end{equation}
Therefore, we can deduce from \fig{fig1} 
how the average vertex order $\overline{o(v)}$ should behave. 
First of all, the average vertex order should be small in the 
weak coupling phase and large in the strong coupling phase. 
As for its behavior for increasing system size, 
in the weak coupling phase 
it should not change much, whereas 
in the strong coupling phase 
it should grow larger and larger.

In \fig{fig3} we show the vertex
order distribution $\rho(n)$ for $\kappa_2$ = 0.0, 1.267
(near the critical point) and $2.0$ with $N_4=32000$.
%
%
For $\kappa_2=2.0$, the distribution damps quite rapidly for 
large vertex order and we have confirmed that it remains almost unchanged 
when we increase the system size. 
For $\kappa_2=0.0$, on the other hand, 
one finds that there is a peak of very large vertex order; 
as large as one third of the total four-simplices.
We have also confirmed that the peak consists of two vertices. 
In \fig{lam0_N4} we show the size dependence of the vertex order distribution 
for $\kappa_2=0.0$.
One finds that the very large vertex order grows linearly 
as one increases $N_4$. 



As was mentioned earlier, in the strong coupling phase, 
one already knows from the behavior of $\langle R_{\av} \rangle$ 
that the average vertex order should grow as the system size is increased. 
However, what we have found by looking at the vertex order distribution 
is that the distribution does not shift to the right uniformly but 
that the two singular vertices move to the right linearly with increasing 
system size, 
leaving the continuum part which, as we also observe, 
moves to the right but much more slowly
than the singular vertices. 
This seems to be a potential obstacle to taking a sensible continuum limit 
from the strong coupling phase. 

As we emphasized earlier, we still have the freedom to modify the action. 
We, therefore, attempt to cure the above disease by adding the 
following term to the action
\footnote{
For some studies on other types of modified action with different 
motivations, 
see Refs. \cite{Amb92a,Brug93a}.
}.
\beq
S_u={u\over 2} \sum_v (o(v)-C)^2.
\label{eq:lambdaTerm}
\eeq
%
We note that, since a change of the constant $C$ 
can be absorbed into a redefinition of $\kappa_2$ and $\kappa_4$, 
we can fix it to an arbitrary value without loss of generality. 
We choose $C=5$, considering that $o(v)\ge 5$ is guaranteed from 
the requirement that the configuration should be a simplicial manifold 
in the strictly mathematical sense.
We have performed simulations with $u=\lamA$, $\lamB$ and $\lamC$.
The vertex order distribution is measured every 100 sweeps and averaged 
over 100 configurations as in the $u=0$ case. 
In order to reduce the fluctuations of the distribution, 
we smear the data over bins of size 5.

In \fig{lfigOvl1} we show the vertex order distribution 
for $u=\lamA$ with $N_4=32000$. 
One sees that there is no singular vertex with very large order
for all $\kappa_2$.
On the other hand, we find that for $u=\lamC$, 
the results resemble the ones for $u=0$. 


Let us next turn to the curvature and the susceptibility.
\fig{lfigR} shows 
$\langle R_{\av}\rangle$ as a function of $\kappa_2$
for $u=\lamA$, $\lamB$ and $\lamC$. We have replotted the data for $u=0$.
For $u=\lamC$,
the curves resemble the $u=0$ case. 
On the other hand, for $u=\lamA$
and 
$\lamB$, 
the curves for $N_4=8000$ and $32000$ coincide 
throughout the $\kappa_2$ region 
examined. 
\fig{lfigRs} shows the susceptibility $\chi_R$ as a function of
$\kappa_2$ for $u=\lamA$,
$\lamB$ and $\lamC$.
Although we see a broad peak even for $u \geq \lamB$, 
it remains unchanged when we increase the system size 
in contrast to the $u=0$ case. 

These results imply that the second order
phase transition which is observed in the $u=0$ case
(\fig{fig2}) disappears for a sufficiently large $u$.
We also observe that the configurations for $u\geq\lamB$ 
have much the same properties as the ones in the weak coupling phase 
of the $u=0$ case, {\it i.e.} no mother universe, low Hausdorff dimension, 
little size dependence in the vertex order distribution, etc.. 
Therefore we claim that in the presence of the additional term, 
with a sufficiently large coefficient in the action, 
the system is in the branched polymer phase for any $\kappa_2$ .  

~

To summarize, we find that, in the strong coupling phase, 
there are two singular vertices with 
very large vertex order, which grows linearly as the system size is increased. 
We consider this to be a potential obstacle to 
taking a sensible continuum limit from the strong coupling phase. 
Adding a term like (\ref{eq:lambdaTerm}) turns out to have as strong an
influence on the system as to erase the phase transition for a sufficiently 
large coefficient.
A local term having such a large influence on the system is in itself 
an interesting fact. 
Also, that we find a branched polymer structure 
with a large average vertex order 
is worth mentioning. 
The fact that the size dependence completely disappears for a sufficiently 
large coefficient of the additional term provides
an intuitive picture of the branched polymer structure : 
the typical configuration is composed of baby universes which are fluctuating 
independently, and increasing the system size merely contributes to increasing 
the number of baby universes without changing the local properties of 
the configuration. 
In order to take a sensible continuum limit, it seems that we need to fine--tune
$u$ or try other types of modification of the action \cite{HIN}. 

~

We would like to thank H. Kawai, T. Yukawa and N. Tsuda for 
stimulating discussion. 
We are also grateful to B. Hanlon for carefully reading the manuscript.



\include{bib}
\include{fig}


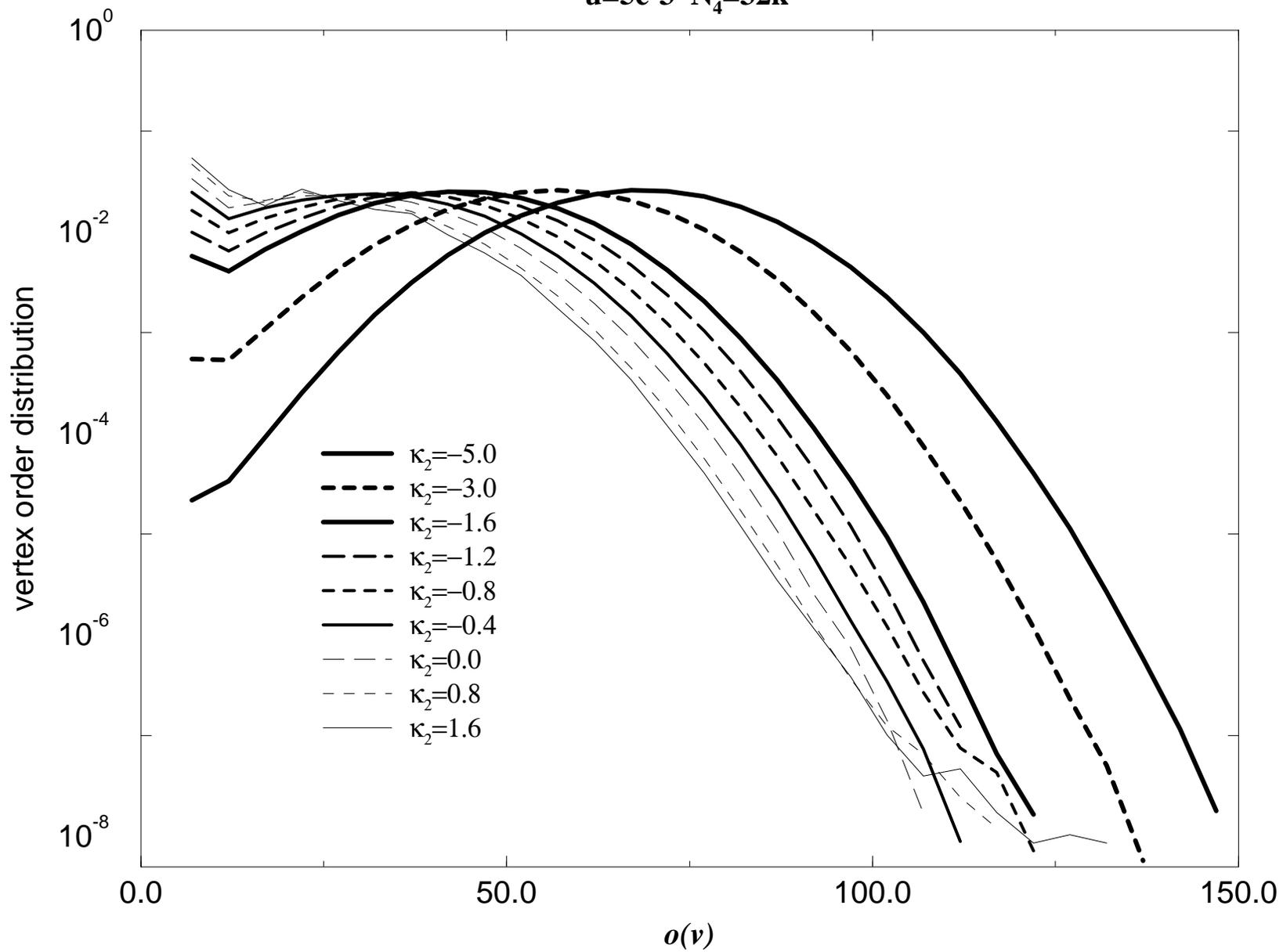
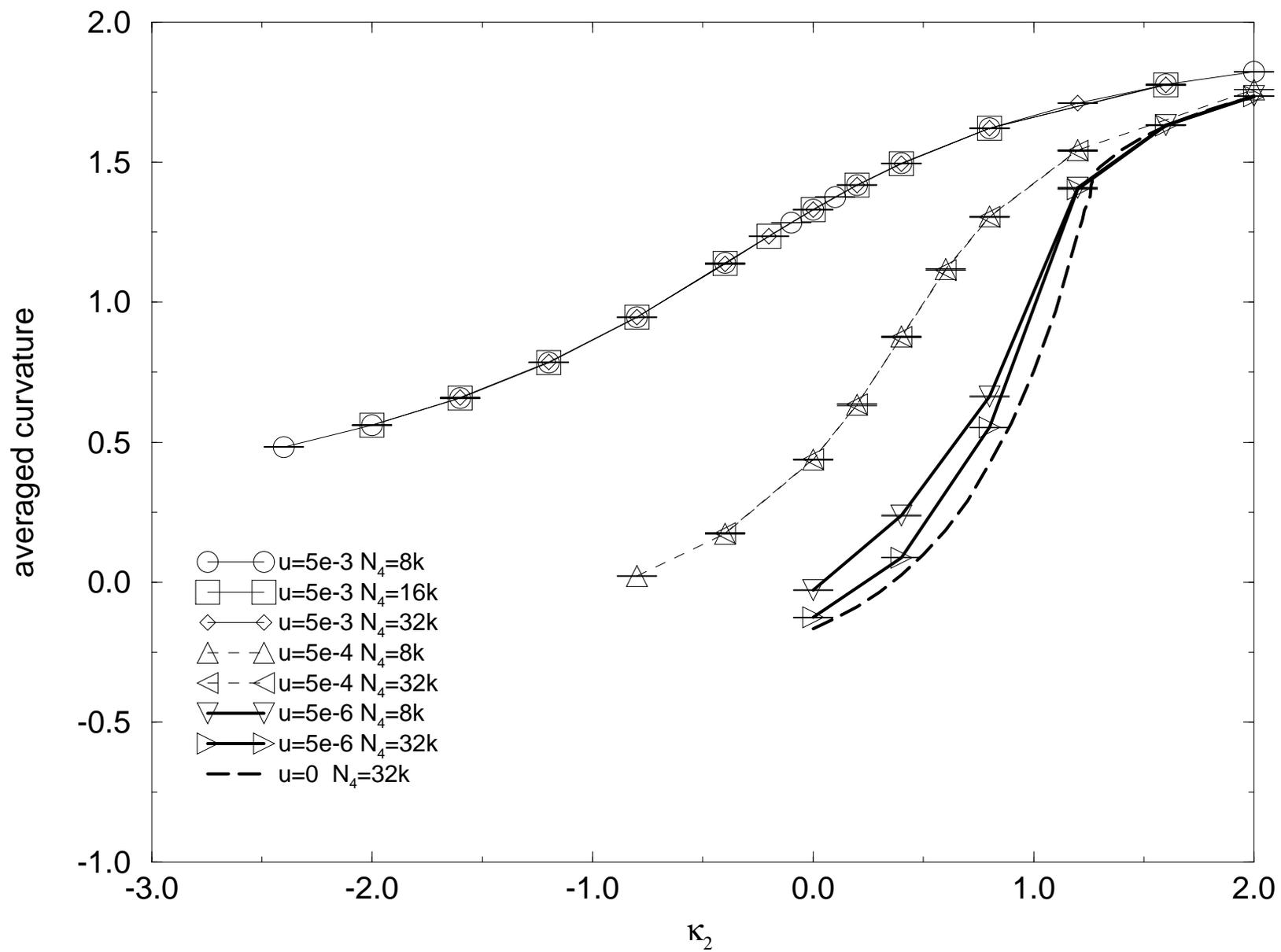
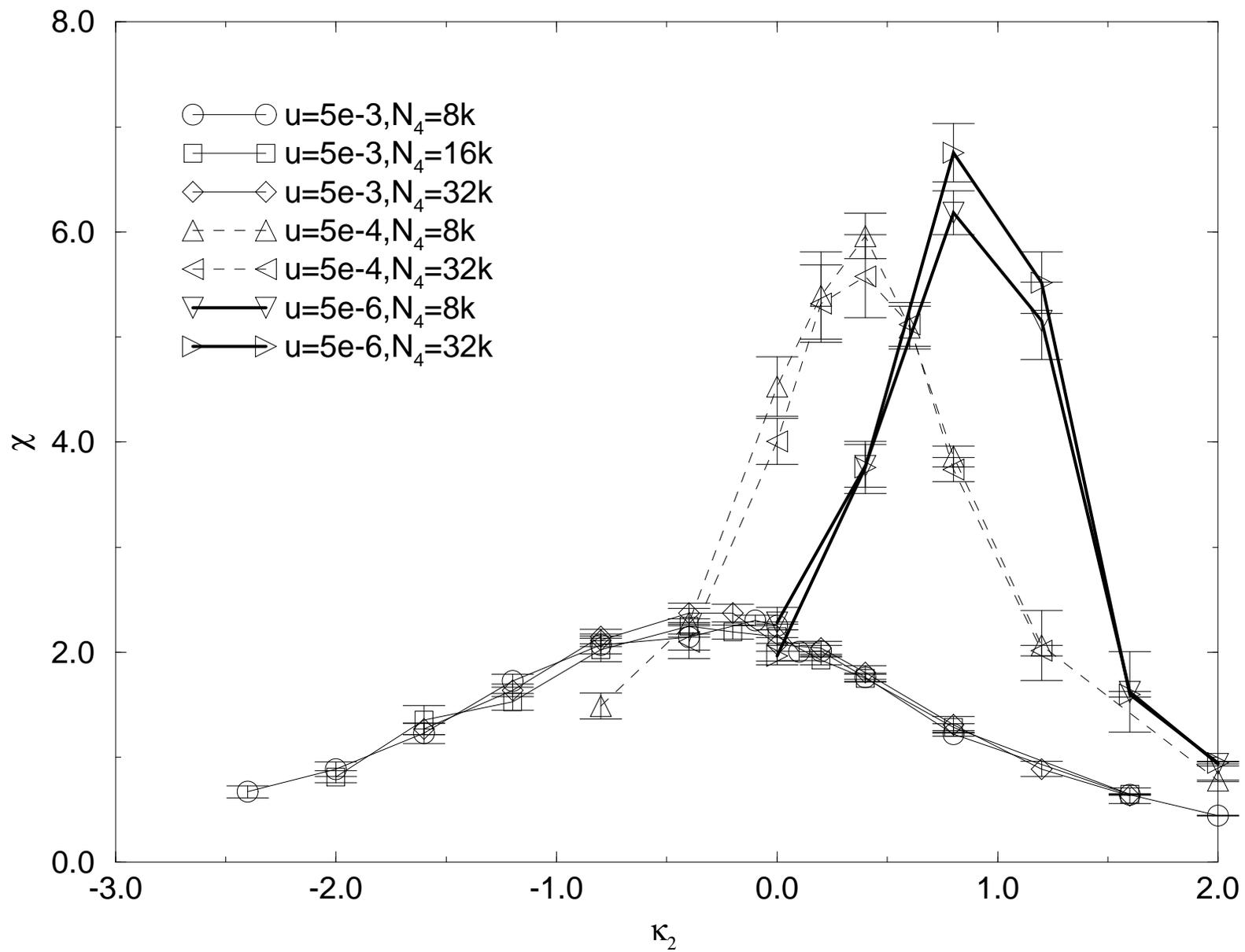
\end{document}

%% file: bib.tex

%% file: fig.tex
\begin{figure}
  \caption{
    The average curvature per unit volume (divided by 
$\alpha=\arccos \left(\frac{1}{4}\right)$) is plotted against 
    $\kappa_2$ for $N_4=8000$ (circles) and $N_4=32000$ (squares).}
  \label{fig:fig1}
\end{figure}
\begin{figure}
  \caption{
    The susceptibility (divided by $\alpha^2$) 
is plotted against $\kappa_2$
    for $N_4=8000$ (circles) and $N_4=32000$ (squares).}
  \label{fig:fig2}
\end{figure}
\begin{figure}
  \caption{
    The vertex order distribution for $\kappa_2=0.0$ (solid curve),
    $1.267$ (dotted curve), and $2.0$ (dashed curve) with
    $N_4=32000$. }
  \label{fig:fig3}
\end{figure}
\begin{figure}
  \caption{
    The vertex order distribution for $\kappa_2=0.0$ with
    $N_4=8000$ (dot-dashed curve),
    $N_4=16000$ (long dashed curve),
    $N_4=32000$ (dashed curve),
    $N_4=64000$ (dotted curve),
and $N_4=128000$ (solid curve).}

  \label{fig:lam0_N4}
\end{figure}

\begin{figure}
\caption{The vertex order distribution for $u=\lamA$
with $N_4=32000$. 
The curves corresponds to $\kappa_2=$-5.0, -3.0, -1.6, -1.2, -0.8, 
-0.4, 0.0, 0.8 and 1.6 respectively. 
}
\label{fig:lfigOvl1}
\end{figure}


\begin{figure}
\caption{The average curvature per unit volume (divided by $\alpha$)
is plotted against $\kappa_2$ 
for $u=\lamA$ (solid curve), 
$\lamB$ (dashed curve), 
$\lamC$ (thick curve) with 
$N_4=$8000 and 32000. 
The result for $u=0$ (thick dashed curve) with $N_4=$32000 is replotted.}
\label{fig:lfigR}
\end{figure}

\begin{figure}
\caption{The susceptibility (divided by $\alpha^2$) 
is plotted against $\kappa_2$ 
for $u=\lamA$ (solid curve), 
$\lamB$ (dashed curve) and $u=\lamC$ (thick solid curve) with 
$N_4=$8000 and 32000.}
\label{fig:lfigRs}
\end{figure}